\documentclass[hidelinks,aps,pra,twocolumn,superscriptaddress,10pt]{revtex4-2} 

\usepackage{amsmath,amssymb}
\usepackage{mathrsfs}
\usepackage{epstopdf}
\usepackage{graphicx}
\usepackage{bm,color,bbm}
\usepackage{natbib}
\usepackage{hyperref}
\usepackage{caption}
\usepackage{fancyhdr}
\usepackage{subcaption}
\newcommand{\nn}{\nonumber}

\begin{document}
\fancyhead[R]{\ifnum\value{page}<2\relax\else\thepage\fi}

\title{Quantifying Tripartite Spatial and Energy-Time Entanglement in Nonlinear Optics}


\author{James Schneeloch}
\email{james.schneeloch.1@us.af.mil}
\affiliation{Air Force Research Laboratory, Information Directorate, Rome, New York, 13441, USA}

\author{Richard J. Birrittella}
\affiliation{Air Force Research Laboratory, Information Directorate, Rome, New York, 13441, USA}

\author{Christopher C. Tison}
\affiliation{Air Force Research Laboratory, Information Directorate, Rome, New York, 13441, USA}

\author{Gregory A. Howland}
\affiliation{Microsystems Engineering, Rochester Institute of Technology, Rochester, New York 14623, USA}
\affiliation{School of Physics and Astronomy, Rochester Institute of Technology, Rochester, New York 14623, USA}

\author{Michael L. Fanto}
\affiliation{Air Force Research Laboratory, Information Directorate, Rome, New York, 13441, USA}

\author{Paul M. Alsing}
\affiliation{Air Force Research Laboratory, Information Directorate, Rome, New York, 13441, USA}


\date{\today}

\begin{abstract}
In this work, we provide a means to quantify genuine tripartite entanglement in arbitrary (pure and mixed) continuous-variable states as measured by the Tripartite Entanglement of formation --- a resource-based measure quantifying genuine multi-partite entanglement in units of elementary Greenberger-Horne-Zeilinger (GHZ) states called gebits. Furthermore, we predict its effectiveness in quantifying the tripartite spatial and energy-time entanglement in photon triplets generated in cascaded spontaneous parametric down-conversion (SPDC), and find that ordinary nonlinear optics can be a substantial resource of tripartite entanglement.
\end{abstract}


\maketitle
\thispagestyle{fancy}
\section{Introduction}
As quantum networking and computing platforms grow more sophisticated, it is more important than ever to develop means of efficiently characterizing the quantum resources present. To that end, many advances have been made over the last few years in quantifying the entanglement present between two groups of increasingly high-dimensional systems \cite{MartinEntRecord,schneeloch2018EntExp}. However, when it comes to quantifying \emph{multi-partite} entanglement in high-dimensional systems, the field remains relatively underdeveloped with notable exceptions \cite{adesso2006continuous,piano2013genuine} using generalizations of the three-tangle \cite{CKWpaper2000}, a monotone based on the residual entanglement \footnote{The residual entanglement between system $A$ and systems $B$ and $C$ is the difference between the entanglement between $A$ and $BC$ jointly, and the sum of the entanglements between $A$ and $B$ and $C$ individually. For pure states, a positive residual entanglement witnesses tripartite entanglement, but it does not faithfully identify all tripartite-entangled states (e.g., the $|W\rangle$ state).}, which identifies most but not all tripartite entangled states. Recently, resource-based entanglement measures that both faithfully identify all multi-partite entangled states, and are additive over copies of the state to be measured have been developed \cite{Szalay_MultiEntMeas,OnoeMultiGaussEnt2020}, but the fundamental challenge of efficiently quantifying genuine multi-partite entanglement in high-dimensional systems remains to be answered \cite{friis2019entanglement}.

As experimental sources of entanglement have a finite preparation uncertainty, strategies toward quantifying multi-partite entanglement must be applicable to mixed as well as pure states. To that end, there do exist multiple witnesses of genuine continuous-variable tripartite entanglement that apply to arbitrary quantum states \cite{LoockFurusawaMultiEnt2003,SaboiaEntRelMultPart2015,TehMultiEpr2014,PhysRevLett.104.210501,universe5100209}, which have been used successfully in experiment \cite{shalm2013three}. In general, witnesses of genuine tripartite entanglement date back as early as 1987 \cite{SvetlichnyIneq1987}, but quantifying more than a nonzero amount of tripartite entanglement present has remained elusive.

Within the last year, the challenge of quantifying genuine tripartite entanglement in mixed states has been answered in part in \cite{Schneeloch_TriEnt}, where correlations between observables of qudits were used to place a lower bound on the tripartite entanglement of formation $E_{3F}$ --- a measure of genuine tripartite entanglement that compares the arbitrary state being measured to a comparable number of three-qubit GHZ states, known as three-party gebits. This strategy to quantify genuine tripartite entanglement was explicitly dependent on the dimension $d$ of the quantum systems, so that adapting it for continuous-variable degrees of freedom remained an open challenge \footnote{In \cite{Schneeloch_TriEnt}, it was shown how genuine tripartite entanglement could be quantified in continuous-variable degrees of freedom, but only on the assumption that the joint state was pure.}. 

In this work, we present a strategy to quantify genuine tripartite entanglement of arbitrary (pure and mixed-state) continuous-variable systems using the correlations naturally present in many of these systems. In particular, we examine the tri-partite spatial and energy-time entanglement present in  cascaded $\chi^{(2)}$ spontaneous parametric down-conversion (SPDC), in which one pump photon is split into two daughter photons, followed by one daughter photon down-converting into two granddaughter photons. The spatial correlations in this system are qualitatively identical to those in $\chi^{(3)}$ SPDC (a single-step photon triplet generation process in nonlinear optics), and generation rates are comparable with one another \cite{PhysRevLett.116.073601,PhysRevLett.118.153602,Krapick:16,moebius2016efficient}, but we focus on cascaded SPDC, as this process is more well-studied \cite{PhysRevA.88.032308}.

\section{Foundations and motivation: Quantifying genuine tripartite continuous-variable entanglement}

Entanglement is defined with respect to separability. Any quantum state $\hat{\rho}$ of three parties $A$, $B$, and $C$, that factors out into a product of states for each party, or any mixture of such factorable states is defined to be separable:
\begin{equation}
\hat{\rho}_{ABC}^{(sep)}=\sum_{i}p_{i}\left(\hat{\rho}_{Ai}\otimes\hat{\rho}_{Bi}\otimes\hat{\rho}_{Ci}\right)
\end{equation}
All other states are entangled.

With more than two parties, there are multiple forms of separability, defining multiple forms of entanglement. For example, states of the form $A\otimes BC$:
\begin{equation}
\hat{\rho}_{A\otimes BC}=\sum_{i}p_{i}\left(\hat{\rho}_{Ai}\otimes\hat{\rho}_{BCi}\right)
\end{equation}
are known as biseparable because they can be expressed as mixtures of states that factor out as a product of two terms --- in this case, one for $A$ and another for the joint state of $BC$. To demonstrate tri-partite entanglement, the state must at least be in no way biseparable, but there is more to it. Proving $ABC$ is genuinely tripartite entangled requires showing not just that the state is outside all three classes of biseparable states (i.e., $A\otimes BC$, $B\otimes AC$ and $C\otimes AB$), but that the state cannot be made out of any arbitrary mixture of states coming from one or more of these classes. This distinction is important because it is possible to combine biseparable states from multiple classes to obtain mixed states that are outside all of these sets. Such fully inseparable states are not genuinely tripartite entangled.

The measure of genuine tripartite entanglement we will be using in this paper is the tripartite entanglement of formation $E_{3F}$, which for parties $A$, $B$, and $C$, is given by:
\begin{equation}
E_{3F}(ABC)=\min_{|\psi\rangle_{i}}\sum_{i}p_{i}\min\{S_{i}(A),S_{i}(B),S_{i}(C)\}
\end{equation}
where the first minimum is taken over all pure state decompositions of $\hat{\rho}_{ABC}$ and the second minimum is of the entanglement entropy over all bipartitions of each constituent pure state in the decomposition. This measure, first discussed in \cite{Szalay_MultiEntMeas}, is a generalization of the regular entanglement of formation, and is: (1) greater than zero if and only if the state is genuinely tripartite entangled; (2) invariant under local unitary transformations; (3) monotonically decreasing under local operations and classical communication (LOCC), and (4) at least additive over tensor products for pure states \footnote{For mixed states, $E_{F}$ only approximately additive. It is sub-additive (i.e., $E_{F}(\hat{\rho}^{\otimes n})\leq n E_{f}(\hat{\rho})$), but is lower-bounded by a fully additive measure that reduces to $E_{F}$ for the pure-state case (i.e., the squashed entanglement).}, so that $m$ copies of a given entangled state will have $m$ times the value of $E_{3F}$ that one copy does. This facilitates side-by side comparisons of multiple low-dimensional entangled states with fewer high-dimensional entangled states. For pure tripartite states, $E_{3F}$ is simply equal to the minimal entropy between subsystems $A$, $B$, and $C$. In \cite{Schneeloch_TriEnt}, we were able to lower-bound $E_{3F}$ using correlations between observables of $d$-dimensional systems, but in this article, we show how one can also do this for continuous-variable (high-dimensional) systems.

To witness genuine tripartite entanglement in both pure and mixed states, one can start with a convex witness of genuine tripartite entanglement for pure states. Here, the convex witness is any convex function $f$ of the quantum state $|\psi\rangle_{ABC}$ such that for some value $\eta$, $f>\eta$ witnesses genuine tripartite entanglement. The convexity allows us to immediately apply these witnesses to mixed states because the average value of a convex function cannot increase under mixing.

Once a convex witness of genuine tripartite entanglement for pure states is found, it is readily adapted to fully general (i.e., mixed) states. Given $f$ is  convex, if $\hat{\rho}$ has a pure state decomposition:
\begin{equation}
\hat{\rho}=\sum_{i}p_{i}|\psi_{i}\rangle\langle\psi_{i}|,
\end{equation}
the witness will obey the inequality:
\begin{equation}
f(\hat{\rho})\leq \sum_{i} p_{i} f(|\psi_{i}\rangle\langle\psi_{i}|)
\end{equation}
Since $f>\eta$ witnesses genuine tripartite entanglement in a pure state, it must also follow that for any mixture of pure states $\hat{\rho}$, that $f(\hat{\rho})>\eta$ witnesses genuine tripartite entanglement as well. This forces at least one element in any pure state decomposition of $\hat{\rho}$ to be genuinely tripartite entangled, which is sufficient to prove genuine tripartite entanglement in the mixed state case. This general strategy of constructing convex witnesses for pure states to adapt them for mixed states has been used to great success to construct multi-partite entanglement witnesses from uncertainty relations in \cite{SaboiaEntRelMultPart2015}.

To \emph{quantify} genuine tripartite entanglement, we use convex entanglement witnesses that bound the quantum conditional entropy. In particular, one can show from what we found in \cite{SchneelochPra2018}, that for Fourier-conjugate position $x$ and momentum $k=p/\hbar$:
\begin{subequations}\label{condEntRel}
\begin{equation}
\log(2\pi)  \!-\! h(x_{A}|x_{B},x_{C}) \!-\! h(k_{A}|k_{B},k_{C})\!\leq \!\!- S(A|BC)
\end{equation}
\begin{equation}
\log(2\pi) \!-\! h(x_{B}|x_{C},x_{A}) \!-\! h(k_{B}|k_{C},k_{A})\!\leq \!\! - S(B|CA)
\end{equation}
\begin{equation}
\log(2\pi) \!-\! h(x_{C}|x_{A},x_{B}) \!-\! h(k_{C}|k_{A},k_{B})\!\leq\!\!  - S(C|AB)
\end{equation}
\end{subequations}
Here, $h(x_{A}|x_{B},x_{C})=h(x_{A},x_{B},x_{C})-h(x_{B},x_{C})$, and  $h(x_{A},x_{B},x_{C})$ is the continuous Shannon entropy 
\cite{Cover2006} of the joint probability density of $x_{A}$, $x_{B}$, and $x_{C}$. In addition, $S(A|BC)=S(ABC)-S(BC)$ is the quantum conditional entropy where for example, $S(ABC)$ is the von Neumann entropy of density matrix $\hat{\rho}_{ABC}$. The left hand sides of \eqref{condEntRel} witness entanglement in their respective bipartitions when they are greater than zero. All logarithms are taken to be base two, since we measure entropy in bits. For a more detailed discussion behind the derivation of these relations, see Appendix \ref{ApA}.

With the preceding three relations \eqref{condEntRel}, we find functions of $x$ and $k$ that bound the left hand side of all three at once. For momentum $k$, we have for the entropy of a linear combination of the three momenta:
\begin{equation}
h(\beta_{A}k_{A}+\beta_{B}k_{B}+\beta_{C}k_{C})\geq
\begin{cases}
h(k_{A}|k_{B},k_{C}) + \log(|\beta_{A}|)\\
h(k_{B}|k_{A},k_{C})+ \log(|\beta_{B}|)\\
h(k_{C}|k_{A},k_{B})+ \log(|\beta_{C}|)
\end{cases}
\end{equation}
where $(\beta_{A},\beta_{B},\beta_{C})$ are real-valued coefficients and $|\cdot|$ denotes absolute value. Similarly for the linear combination of positions, we have the relation:
\begin{equation}
h\left(\eta_{A}x_{A}+\eta_{B}x_{B}+\eta_{C}x_{C}\right)\geq
\begin{cases}
h(x_{A}|x_{B},x_{C})+\log(|\eta_{A}|)\\
h(x_{B}|x_{A},x_{C})+\log(|\eta_{B}|)\\
h(x_{C}|x_{A},x_{B})+\log(|\eta_{C}|)
\end{cases}
\end{equation}
See proof in Appendix \ref{ApC}.
Together, these allow us to consolidate these separate bounds \eqref{condEntRel} into one:
\begin{align}
h\left(\eta_{A}x_{A}\!+\!\eta_{B}x_{B}\!+\!\eta_{C}x_{C}\right)&\!+\!h(\beta_{A}k_{A}\!+\!\beta_{B}k_{B}\!+\!\beta_{C}k_{C})\nn\\
&\!\!\!\!\!\!\!\!\!\!\!\!\!\!\!\!\!\!\!\!\!\!\!\!\geq\log(2\pi)+\log\left(\min_{i}\{|\eta_{i}||\beta_{i}|\}\right)\nn\\
+\max&\{S(A|BC),S(B|AC),S(C|AB)\}
\end{align}
To obtain a bound for the tripartite entanglement of formation for pure states, we use the relation for pure states that $-S(A|BC)=S(A)$, and re-arrange the previous inequality to obtain:
\begin{align}
\min\{&\!S(\!A),\!S(\!B),\!S(\!C)\!\}\!\geq \!\log(2\pi\!|\bar{\eta}||\bar{\beta}|)\nn\\
&\!\!\!\!\!\!-h\left(\eta_{A}x_{A}\!+\!\eta_{B}x_{B}\!+\!\eta_{C}x_{C}\right)-h(\beta_{A}k_{A}\!+\!\beta_{B}k_{B}\!+\!\beta_{C}k_{C})
\end{align}
where $|\bar{\eta}||\bar{\beta}|=\min_{i}\{|\eta_{i}||\beta_{i}|\}$.This relation is true for every element in the pure state decomposition of $\hat{\rho}_{ABC}$, and therefore for any mixture. Then, because this relation must hold, even if we choose the pure state decomposition that minimizes the left hand side of this relation, we have our bound for the tripartite entanglement of formation $E_{3F}$:
\begin{align}\label{BigResult}
E_{3F}&(ABC)\!\geq \!\log(2\pi\!|\bar{\eta}||\bar{\beta}|)\\
&\!\!\!\!\!\!-h\left(\eta_{A}x_{A}\!+\!\eta_{B}x_{B}\!+\!\eta_{C}x_{C}\right)-h(\beta_{A}k_{A}\!+\!\beta_{B}k_{B}\!+\!\beta_{C}k_{C})\nn
\end{align}
With this bound, we can place a conservative lower limit to $E_{3F}$ on continuous-variable systems where direct calculation is generally intractable even with full knowledge of the state. Moreover, this relation can be adapted into one using standard deviations $\sigma$ or variances $\sigma^{2}$ instead of entropies (as was accomplished previously for bipartite entanglement in \cite{SchneelochPra2018}), because the Gaussian distribution is the maximum entropy distribution for a fixed variance:
\begin{align}
E_{3F}(ABC)&\geq -\log\Bigg(\frac{e}{|\bar{\eta}||\bar{\beta}|} \sigma\left(\eta_{A}x_{A}\!+\!\eta_{B}x_{B}\!+\!\eta_{C}x_{C}\right)\nn\\
&\cdot \sigma\left(\beta_{A}k_{A}\!+\!\beta_{B}k_{B}\!+\!\beta_{C}k_{C}\right)\Bigg)
\end{align}
Resource measures of multi-partite entanglement are still relatively underdeveloped, but we expect that tools such as these will spur new growth in the field.

\subsubsection{Generality of application}
While the entanglement bound \eqref{BigResult} covers arbitrary linear combinations of positions or momenta, we will focus for the rest of the paper on the relation adapted for correlations seen in simple nonlinear-optical sources of spatially entangled photon triplets. In particular, we consider the case of a pump photon being converted into an entangled photon triplet through nonlinear-optical processes. Conservation of momentum implies that the uncertainty in the sum of the triplet's momenta $k_{A}+k_{B}+k_{C}$ is bounded by the uncertainty in the pump momentum, which may be made arbitrarily narrow. If we consider that these photon triplets may arise from a common birthplace, then we may expect the mean squared distance between one of the photons $x_{A}$, and their common centroid $(x_{A}+x_{B}+x_{C})/3$ to be small as well, so that the uncertainty in the linear combination $x_{A}-(x_{B}+x_{C})/2$ could also be arbitrarily narrow. In cascaded SPDC, we expect the uncertainty in $x_{A}-(x_{B}+x_{C})/2$ to be narrow as well. In this case, there are two birthplaces; one between $x_{A}$ and the centroid $(x_{B}+x_{C})/2$ for the first down-conversion event, and one between $x_{B}$ and $x_{C}$ for the second event. The principal distance here is between $x_{A}$ and the centroid $(x_{A}+(x_{B}+x_{C})/2)/2$ which gives, up to a constant factor the quantity $x_{A}-(x_{B}+x_{C})/2$ just as before. For the rest of the paper, we will be considering the form of our relation where $(\eta_{A},\eta_{B},\eta_{C})=(1,-1/2,-1/2)$ and $(\beta_{A},\beta_{B},\beta_{C})=(1,1,1)$.

Although we derived our tripartite entanglement bound \eqref{BigResult} using Fourier-conjugate position $x$ and momentum $k=p/\hbar$, any pair of Fourier-conjugate variables will apply, including time $t$ and (angular) frequency $\omega$, or functions of conjugate field quadratures in quantum optics as studied in \cite{olsen2018tripartite,gonzalez2018continuous}.

Indeed, in \cite{shalm2013three}, Shalm \emph{et~al} demonstrated tri-partite energy-time entanglement between photon triplets generated in cascaded SPDC, subject to the assumption that $\sigma(\omega_{1}+\omega_{2}+\omega_{3})=\sigma(\omega_{p})$. Their source generates photon triplet detection events at a rate of approximately seven per hour of the course of three days, but using their maximum time uncertainty $\sigma(t_{2}-t_{1})=3.7\times10^{-10}$ s as an approximation toward the uncertainty $\sigma(t_{A}\!\!-\!\!\frac{t_{B}\!+\!t_{C}}{2})$, and their pump uncertainty $\sigma(\omega_{p})=3.77\times10^{7}/$s, one could verify as much as $3.72$ gebits of tripartite energy-time entanglement, which is already more entanglement than an $11$-qubit or $2200$-dimensional state can support.

\section{Effectiveness in cascaded SPDC}

\begin{figure}[t]
\includegraphics[width=0.9\columnwidth]{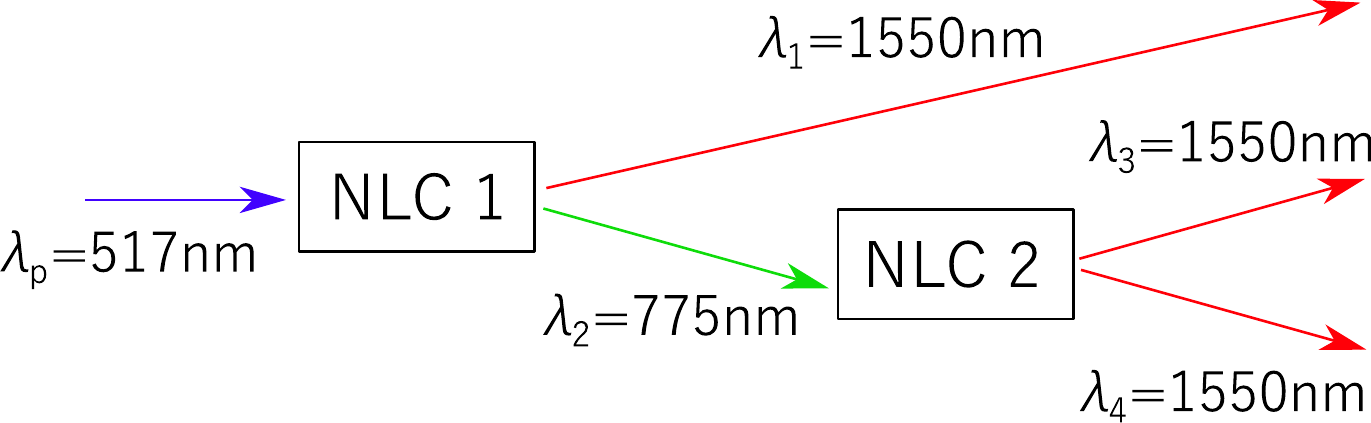}
\caption{Basic diagram of degenerate-cascaded SPDC using $517$nm pump light to produce triplets at $1550$nm. First, light at $\lambda_{p}$ is split into wavelengths $\lambda_{1}$ and $\lambda_{2}$. The light at $\lambda_{2}$ is then split into light at $\lambda_{3}$ and $\lambda_{4}$.}
\end{figure}

Having developed our quantitative bound for tripartite entanglement, we now test its effectiveness for a realistic source of tripartite entanglement. In this work, we consider a source of spatially entangled photon triplets generated in degenerate cascaded spontaneous parametric down-conversion (See Fig.~1 for basic diagram). The source would be a pump laser at frequency $\omega_{p}$ interacting with two $\chi^{(2)}$-nonlinear crystals each of length $L_{z}$. The first crystal would be phase-matched to produce signal/idler photon pairs at frequencies $2\omega_{p}/3$ and $\omega_{p}/3$, respectively. The signal photons would then be directed toward the second crystal chosen to be phase matched for degenerate SPDC taking idler photons at $2\omega_{p}/3$ and producing photon pairs at $\omega_{p}/3$.

As shown in Appendix \ref{ApD}, the transverse spatial amplitude of the photon triplets generated in this process is of the form:
\begin{equation}
\psi(\vec{q}_{1},\vec{q}_{3},\vec{q}_{4})=\mathcal{N}\alpha_{qp}(\vec{q}_{1}+\vec{q}_{3}+\vec{q}_{4})\text{sinc}\left(\frac{\Delta k_{z} L_{z}}{2}\right)
\end{equation}
where here:
\begin{equation}
\Delta k_{z} \approx \frac{3}{2|k_{\tilde{p}}|}\left(|\vec{q}_{1}+\vec{q}_{3}|^{2}+|\vec{q}_{1}+\vec{q}_{4}|^{2}+|\vec{q}_{3}+\vec{q}_{4}|^{2}\right).
\end{equation}
In these expressions: $\alpha_{qp}$ is the transverse momentum amplitude of the pump field where $\vec{q}_{p}$ is set equal to $(\vec{q}_{1}+\vec{q}_{3}+\vec{q}_{4})$ by transverse momentum conservation; $\vec{q}_{1}$ is the projection onto the transverse plane of the momentum $\vec{k}_{1}$ of the lower energy idler photon exiting the first crystal; $\vec{q}_{3}$ and $\vec{q}_{4}$ are the corresponding transverse-projected momenta of the photons created in the second crystal; $k_{\tilde{p}}=k_{p} + k_{\Lambda_{1}} + k_{\Lambda_{2}}$, where $k_{\Lambda_{1}}$ and $k_{\Lambda_{2}}$ are the poling momenta of the first and second crystals; and $\mathcal{N}$ is a normalization constant. If no periodic poling or quasi-phase matching is employed to achieve these processes, then $(k_{\Lambda_{1}} = k_{\Lambda_{2}}=0)$.

To simplify notation, we will let $k_{1}$ refer to the first transverse component of $\vec{k}_{1}$, and define $k_{3}$ and $k_{4}$ similarly. Since for small arguments of the Sinc function, $\text{sinc}(x^{2}+y^{2})\approx \text{sinc}(x^{2})\text{sinc}(y^{2})$, we have as our model for the triphoton wavefunction (for one spatial component):
\begin{align}
\psi(k_{1}&,k_{3},k_{4})\approx\mathcal{N}\alpha_{p}(k_{1}+k_{3}+k_{4})\times\nn\\
&\times\text{sinc}\left(\!\frac{3 L_{z}}{4 k_{\tilde{p}}}\!\left(\!(k_{3}\!+\!k_{4})^{2} \!\!+\!\!(k_{1}\!+\!k_{3})^{2}\!\!+\!\!(k_{1}\!+\!k_{4})^{2}\right)\!\!\right)
\end{align}
This function is symmetric under permutations of $k_{1}$, $k_{3}$ and $k_{4}$. Moreover, the argument of the sinc function is a quadratic form, so we can simplify it dramatically by transforming to a new basis of coordinates. Taking this, together with the Gaussian approximation of the sinc function in \cite{Schneeloch_SPDC_Reference_2016}, we obtain a simple triple-Gaussian wavefunction for the photon triplets:
\begin{equation}
\psi(k_{u},k_{v},k_{w})=\mathcal{N} e^{-\left(\frac{32a}{9} + 3\sigma_{p}^{2}\right)k_{u}^{2}}e^{-\frac{8a}{9}k_{v}^{2}}e^{-\frac{8a}{9}k_{w}^{2}}
\end{equation}
such that $a=\frac{3 L_{z}}{4 k_{\tilde{p}}}$, and:
\begin{align}
k_{u}&=\frac{1}{\sqrt{3}}(k_{1}+k_{3}+k_{4})\\
k_{v}&=\frac{2}{\sqrt{6}}\left(-k_{1}+\frac{k_{3}+k_{4}}{2}\right)\\
k_{w}&=\frac{1}{\sqrt{2}}(k_{3}-k_{4})
\end{align}
and $\sigma_{p}$ is the ordinary pump beam radius (i.e., one quarter of the $1/e^{2}$ beam diameter). Remarkably, our tripartite entanglement bound can also be expressed in terms of these rotated coordinates:
\begin{equation}
E_{3F}(ABC)\geq\log(\pi)-h\left(\frac{\sqrt{6}}{2}x_{v}\right)-h\left(k_{u}\sqrt{3}\right)
\end{equation}
which is further simplified by virtue of its being a Gaussian distribution to:
\begin{equation}
E_{3F}(ABC)\geq-\log\left(3\sqrt{2} e\right)-\frac{1}{2}\log(\sigma_{ku}^{2}\sigma_{xv}^{2})
\end{equation}
with variances:
\begin{equation}\label{sigmas}
\sigma_{ku}^{2}=\frac{1}{4\left(\frac{32a}{9} + 3\sigma_{p}^{2}\right)}\qquad:\qquad \sigma_{xv}^{2}=\frac{8 a}{9}
\end{equation}
and we obtain the final result:
\begin{equation}\label{Result2}
E_{3F}(ABC)\geq\frac{1}{2}\log\left(16 + \frac{18 \sigma_{p}^{2} k_{\tilde{p}}}{L_{z}}\right)-\log(3\sqrt{2}e).
\end{equation}

\begin{figure}[t]
\includegraphics[width=0.9\columnwidth]{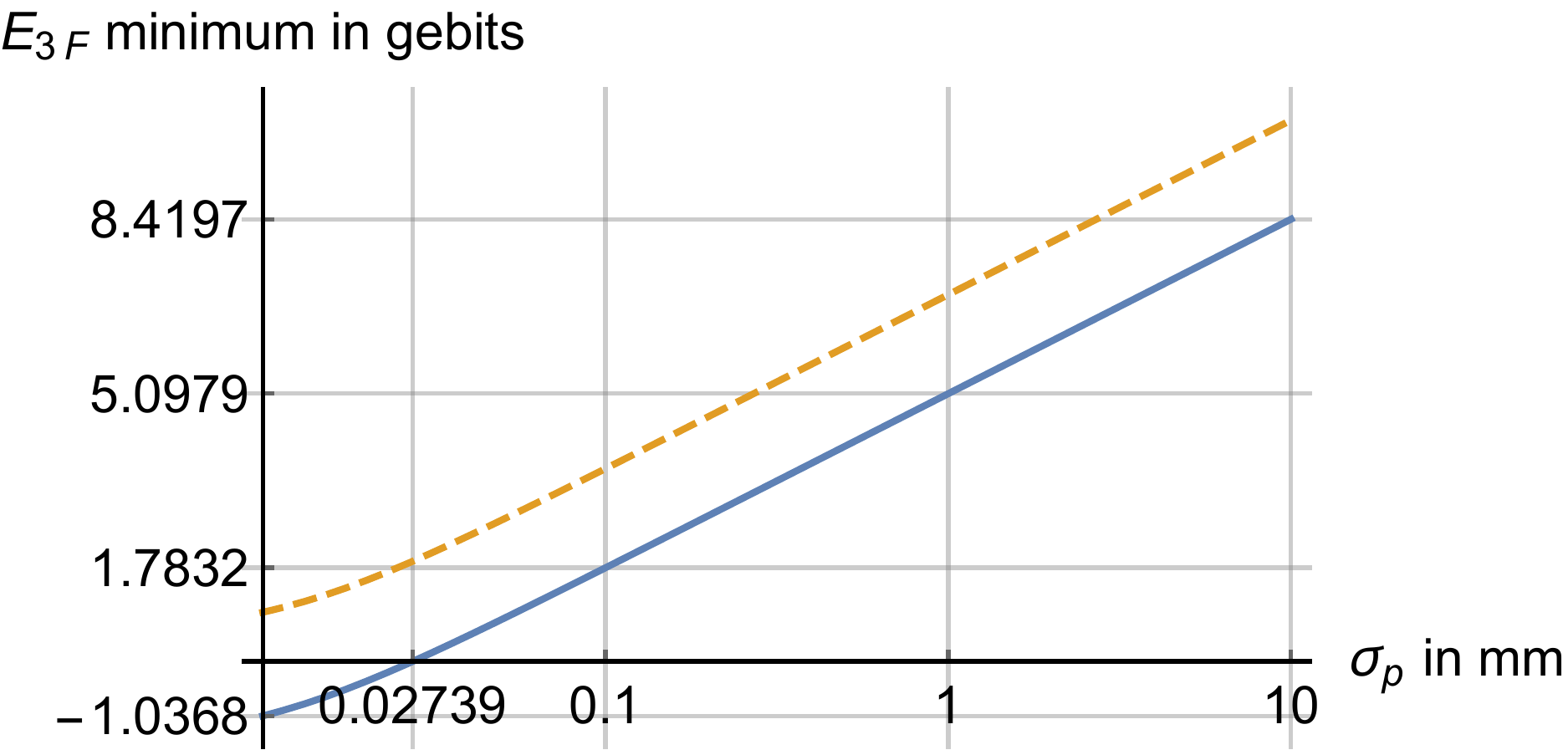}
\caption{Plot of the lower limit to $E_{3F}$ in our considered experimental setup of a pair of PPLN crystals of length $L_{z}=3$mm, as a function of the pump beam radius $\sigma_{p}$ in millimeters. The lower blue curve gives the amount of tripartite entanglement witnessed by our correlation techniques. The upper dashed yellow curve gives the exact value for $E_{3F}$ for the triple-Gaussian wavefunction. In the limit of high correlation, the difference between these two curves raplidly approaches a constant of $2\log_{2}(e)-1$ or about $1.88$ gebits.}
\end{figure}

\subsection{Predicted tripartite entanglement for reasonable experimental parameters}
Let us consider two crystals of periodically poled lithium niobate PPLN whose poling is chosen to be phase matched to the appropriate down-conversion process. At the pump wavelength of $516.67$nm, the index of refraction $n_{p}$ is approximately $2.240$, and the poling periods will have to be about $8.84\mu$m and $18.99\mu$m for each crystal, respectively. From this, the effective pump momentum $k_{\tilde{p}}$ is approximately $2.60\times10^{7}/m$. Bulk crystals come in a variety of lengths, but let us assume $L_{z}=3$mm. The only remaining parameter to fix is the Gaussian beam radius $\sigma_{p}$.

In Fig.~2, we have plotted our lower bound for $E_{3F}$ as a function of $\sigma_{p}$ and find that this source has potentially a substantial amount of entanglement.  In particular, we find for these experimental parameters that a modest beam radius of $1$ mm will generate in excess of five gebits of tripartite entanglement in each transverse dimension, giving us in excess of 10 gebits in total. As a basis of comparison, ten gebits of tripartite entanglement is the maximum amount of tripartite entanglement that a 30-qubit or billion-dimensional state can support! 

\begin{figure*}[t]
\includegraphics[width=0.8\textwidth]{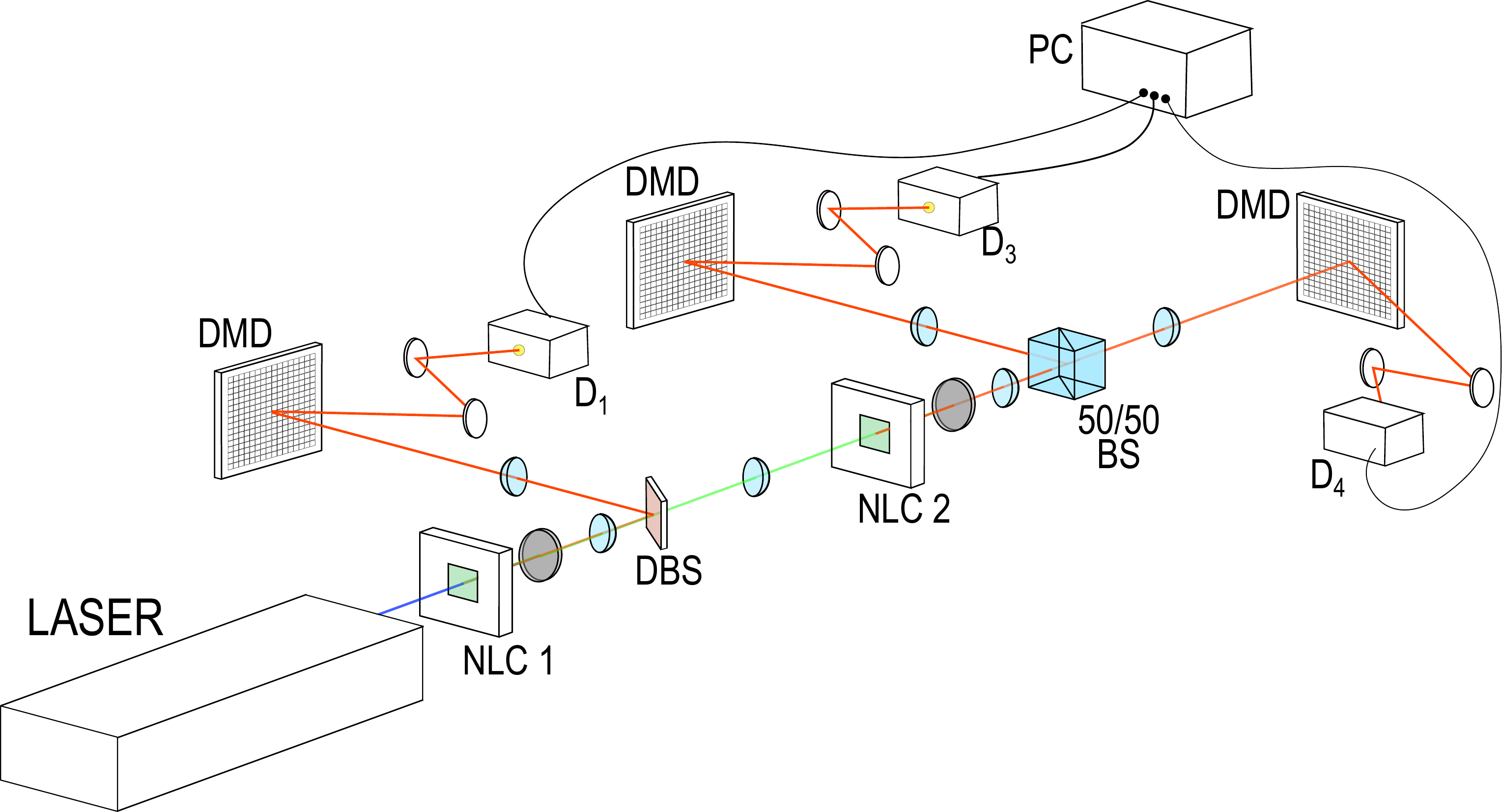}
\caption{Diagram of considered experimental setup to measure position correlations between photon triplets generated in cascaded SPDC. Lenses (in blue) are used both to image the plane of the first nonlinear crystal NLC1 onto a digital micromirror device (DMD) and the plane of the second nonlinear crystal (NLC2), and to image the plane of NLC2 onto the final pair of DMD arrays. The three detectors $D_{1}$, $D_{3}$ and $D_{4}$ are connected to a photon correlator (PC) to record triplet coincidence count rates for each setting of the three DMD arrays, and from this build the joint position probability distribution of the photon triplets from which correlations are obtained.}
\end{figure*}

In order to gauge the effectiveness of our technique at quantifying tripartite entanglement, we need to compare the entanglement we can quantify to the total entanglement present in the triple-Gaussian state. Luckily, the triple-Gaussian wavefunction is simple enough to find its Schmidt decomposition using properties of the double-Gaussian wavefunction \cite{LawEberly2004} (see Appendix \ref{ApE} for details). For both the double-Gaussian wavefunction of two parties, and the triple-Gaussian wavefunction of three parties, the reduced density operator $\hat{\rho}_{A}$ has an identical form. The marginal eigenvalues of $\hat{\rho}_{A}$ are then determined, and from them, the von Neumann entropy $S(A)$ as well. Finally, because the triple-Gaussian wavefunction corresponds to a pure state, and because it is symmetric under permutations of parties, the exact tripartite entanglement of formation is given as $E_{3F}(ABC)=S(A)$, the von Neumann entropy of system $A$. In Fig.~2, we plot both the exact value for $E_{3F}$ for the triple-Gaussian wavefunction along with our lower bound to it \eqref{Result2} based on measured correlations \eqref{BigResult}, and find the gap between the witnessed and total tripartite entanglement rapidly approaches a constant of about $1.88$ gebits in the limit of high correlation. 

As the  pump beam radius $\sigma_{p}$ grows wider (alternatively $\sigma_{xu}$), the corresponding uncertainty in the transverse momentum $\sigma_{ku}$ grows smaller while the other principal variances $\sigma_{kv}$ and $\sigma_{kw}$ remain constant. This results in the momentum distribution becoming more correlated to a flat plane, and the position distribution more correlated to a single line. That both the position and momentum distributions for three particles would be strongly correlated to single lines is actually forbidden by the uncertainty principle \cite{Schneeloch_TriEnt}, a surprising result since this restriction on correlations does not exist between two particles. Even so, the tripartite correlations discussed in this proposed experiment are optimal in that they approach a continuous-variable analogue of the GHZ state.

\subsection{Considered Experimental Setup}
In \cite{schneeloch2018EntExp}, we showed how one can use a simple experimental setup along with techniques in information theory to adaptively sample the transverse position and transverse momentum correlations in such a way that a very small number of measurements (relative to the state space) can faithfully extract those correlations without over-estimating the entanglement present. In this subsection, we discuss how to do the same for the tripartite entanglement generated in our considered setup, referring to Fig.~3 for details.

To begin with, we consider a $517$ nm pump laser incident on a nonlinear crystal (labeled NLC 1) quasi-phase-matched for type-I collinear nondegenerate SPDC with signal light centered at $\lambda_{1}=775$ nm and idler light centered at $\lambda_{2}=1550$ nm, respectively. After passing through a pump removal filter, the light would be split by a dichroic beamsplitter (labeled DBS) that transmits 775nm light and reflects 1550nm light. The length of the reflected 1550nm path following the DBS would terminate with a digital micromirror device (DMD) array whose pixels are computer-controlled to reflect the down-converted light toward or away from a photon-counting detector. This arm would be fitted with optics to either image the plane of NLC 1, or its Fourier transform, giving us access to either the position or momentum statistics of that idler photon, respectively. The transmitted arm following the DBS would terminate with the second crystal (labeled NLC 2), and be fitted with 4F imaging optics to preserve the amplitude and phase of the signal photon.

Next, the signal light at $\lambda_{2}=775$nm would pass through the second nonlinear crystal phase-matched and periodically poled for type-0 degenerate collinear SPDC from $775$ nm converting into photon pairs centered at $\lambda_{3}=\lambda_{4}=1550$ nm. Following NLC2, the residual $775$ nm light would be filtered out, and the $1550$ nm photon pairs would be split by a $50/50$ beamsplitter. Each arm of this wing of the experiment would terminate in a DMD array with optics to image either the field at the plane at NLC 2 onto them, or to image its Fourier transform, thus allowing us access to the position and momentum statistics of this pair as well.

By connecting all three photon detectors to a multi-channel photon correlator (PC), we can record the triplet photon coincidences that occur. By correlating the triplet coincidence count rate to the settings on each DMD array, we can build up a tri-partite joint probability distribution for the positions or the momenta of the photon triplets. 

If we were to build up the joint position and momentum statistics one pixel triplet at a time, acquiring the entire distribution at decent resolution would rapidly become intractable. Indeed, current triplet generation rates are improving for both $\chi^{(2)}$ and $\chi^{(3)}$ processes \cite{hamel2014direct,Krapick:16,moebius2016efficient}, but still are well below one triplet per second per mW of pump power. However, using multi-resolution sampling techniques employed in \cite{schneeloch2018EntExp} can solve this scaling issue as more efficient sources of entangled photon triplets continue to be developed. By sampling first at the lowest possible resolution (i.e., $2\times 2 \times 2$), and next subsampling only in regions with a significant triplet coincidence rate, and iteratively subsampling in the brightest areas of those sub-regions, one can obtain a coarse-grained approximation to these distributions that will never overestimate the entanglement present, and requires a minuscule fraction of the total number of measurements that would be required otherwise. Indeed, in \cite{schneeloch2018EntExp}, these multi-resolution techniques improved the required acquisition time by at least a factor of $10^{7}$, and this advantage will only be more dramatic in the tripartite case, due to the higher dimensional space in which these sparse correlations reside.

\section{Discussion: Resource measure challenges and future applications}
The major issue with generating tripartite spatial entanglement via cascaded SPDC is dealing with a low generation rate. Even with an optimistic free-space generation rate of $10^{8}$ photon pairs per second per mW of pump power, this implies about $1$ in $2.9\times 10^{7}$ pump photons get converted in each crystal, so that only about $1$ in $1.5 \times 10^{15}$ pump photons end up yielding triplets. For $1$ mW of pump power at $517$ nm, this would imply a total generation rate of about $2.6$ triplets per second. Incorporating reasonable sources of loss reduces this rate by an order of magnitude, which puts it on par with the recent demonstration of a triplet generation rate of $12.4$ triplets per minute \cite{hamel2014direct}. Even with the highly efficient structured sensing approach discussed here and in \cite{schneeloch2018EntExp}, the total acquisition time required would make acquiring these spatial correlations at high resolution and statistical significance beyond current capabilities. However, with the recent advent of spatially-resolving photon detectors (e.g., SPAD arrays), demonstrating tripartite spatial entanglement, even at these low intensities becomes possible. With SPAD arrays one can acquire the maximum information from every photon triplet since the position that they strike each detector would be immediately stored, and a usable tripartite position probability distribution can be acquired with a comparably small number of photons.

\section{Conclusion}
The utilization and characterization of multi-partite entanglement is rapidly developing, even while fundamental questions remain to be answered. Here we have presented the first (to our knowledge) technique to quantify genuine tripartite entanglement in continuous-variable systems without the restriction to pure states, enabling us to employ these techniques experimentally. Moreover, we explore a natural source of these tripartite correlations in cascaded spontaneous parametric down-conversion, and find that for reasonable experimental parameters, there is already more tripartite entanglement present than 2-3 dozen qubits can support. On top of this, we were able to gauge the effectiveness of our technique because the high symmetry of the triple-Gaussian wavefunction allowed an explicit calculation of its tripartite entanglement of formation. With current sources of high-dimensional tripartite entanglement owing their strength to the correlations arising from conserved quantities in their interaction,  measurement techniques capitalizing on those correlations are highly efficient. Moreover, using entropy-based tools will allow us to efficiently acquire these correlations at variable resolution without ever over-estimating the entanglement present, as was accomplished for two-party entanglement in \cite{schneeloch2018EntExp}.

\begin{acknowledgments}
We gratefully acknowledge support from the Air Force Office of Scientific Research LRIR 18RICOR028, as well as insightful discussions with Dr. A. Matthew Smith, Dr. H Shelton Jacinto, and an insightful anonymous referee who improved the generality of our results. 

The views expressed are those of the authors and do not reflect the official guidance or position of the United States Government, the Department of Defense or of the United States Air Force. The appearance of external hyperlinks does not constitute endorsement by the United States Department of Defense (DoD) of the linked websites, or of the information, products, or services contained therein. The DoD does not exercise any editorial, security, or other control over the information you may find at these locations.
\end{acknowledgments}

\bibliography{EPRbib16}

\newpage
\appendix

\section{Proof for tripartite entropic uncertainty relation}\label{ApA}
In \cite{SchneelochPra2018}, we proved the relation
\begin{equation}\label{lilrel}
h(x_{A}|x_{B}) + h(k_{A}|k_{B}) \geq \log(2\pi) + S(A|B)
\end{equation}
starting from the entropic uncertainty principle in the presence of quantum memory \cite{Berta2010}, where for a pair of $N$-dimensional observables $\hat{Q}$ and $\hat{R}$, we have:
\begin{equation}\label{BertaRel}
H(Q_{A}|B)+H(R_{A}|B)\geq \log(\Omega_{QR}) + S(A|B)
\end{equation}
where $\Omega_{QR}$ is an uncertainty bound which approaches $N$ in the limit that $\hat{Q}$ and $\hat{R}$ are mututally unbiased. Here, $H(Q_{A}|B)$ is a quantum conditional entropy $S(A|B)$ where observable $\hat{Q}_{A}$ has been measured. This measurement acts as a sum of projectors $|q_{Ai}\rangle\langle q_{Ai}|\otimes I_{B}$, leaving system $A$ in a mixed state of eigenstates of $\hat{Q}_{A}$, but with $B$ unperturbed.

From the fact that quantum conditional entropy can be expressed as relative entropy, we can use the monotonicity of relative entropy to say that any subsequent measurement of system $B$ cannot decrease the left hand side of \eqref{BertaRel}, which gives for arbitrary observables $\hat{V}_{B}$ and $\hat{W}_{B}$, the relation:
\begin{equation}\label{BigBertaRel}
H(Q_{A}|V_{B})+H(R_{A}|W_{B})\geq \log(\Omega_{QR}) + S(A|B)
\end{equation}
Here, there is no assumption that the dimension of system $A$ be the same as system $B$ or that observables $\hat{V}_{B}$ and $\hat{W}_{B}$ are in any way related to $\hat{Q}_{A}$ and $\hat{R}_{A}$. To obtain the relation \eqref{lilrel} from \eqref{BigBertaRel}, we selected $\hat{Q}_{A}$ and $\hat{R}_{A}$ to be a pair of observables related by a quantum Fourier transform, took a pair of continuum limits, and noted that the continuous entropy of the discrete approximation of a random variable is an upper bound to the true continuous entropy of the variable itself. For a full discussion, see \cite{SchneelochPra2018}.

To prove the tripartite version of \eqref{lilrel}:
\begin{equation}\label{bigrel}
h(x_{A}|x_{B},x_{C}) + h(k_{A}|k_{B},k_{C}) \geq \log(2\pi) + S(A|BC)
\end{equation}
we start by taking system $B$ in \eqref{lilrel} to be the joint system $BC$ and let the observables $\hat{V}_{B}$ and $\hat{W}_{B}$ be joint measurements of local observables $(\hat{V}_{B},\hat{V}_{C})$ and $(\hat{W}_{B},\hat{W}_{C})$, of system $BC$. From this, one can take precisely the same steps to derive \eqref{bigrel} as were used to derive \eqref{lilrel} in \cite{SchneelochPra2018}.

\section{Proof of position and momentum entropic bounds}\label{ApC}
In Section II, we gave the following relations for continuous entropy:
\begin{equation}\label{momrel}
h(\beta_{A}k_{A}+\beta_{B}k_{B}+\beta_{C}k_{C})\geq
\begin{cases}
h(k_{A}|k_{B},k_{C}) + \log(|\beta_{A}|)\\
h(k_{B}|k_{A},k_{C})+ \log(|\beta_{B}|)\\
h(k_{C}|k_{A},k_{B})+ \log(|\beta_{C}|)
\end{cases}
\end{equation}
and for position, the relation:
\begin{equation}\label{posrel}
h\left(\eta_{A}x_{A}+\eta_{B}x_{B}+\eta_{C}x_{C}\right)\geq
\begin{cases}
h(x_{A}|x_{B},x_{C})+\log(|\eta_{A}|)\\
h(x_{B}|x_{A},x_{C})+\log(|\eta_{B}|)\\
h(x_{C}|x_{A},x_{B})+\log(|\eta_{C}|)
\end{cases}
\end{equation}

To prove these relations, we will use the following five properties of continuous entropy:

(a) The scaling law for continuous entropy:
\begin{equation}
h(ax) = h(x)+\log(|a|)
\end{equation}
which also implies:
(b) the continuous entropy is constant under reflection:
\begin{equation}
h(x)=h(-x)
\end{equation}
(c) Conditioning cannot increase continuous entropy:
\begin{equation}
h(x)\geq h(x|y)\geq h(x|y,z)\geq ...
\end{equation}
(d) Shifting by a conditioned variable cannot change entropy
\begin{equation}
h(x\pm y|y)=h(x|y)
\end{equation}
(e) Conditioning on functions of already conditioned variables cannot change the entropy:
\begin{equation}
h(x|y,z)=h(x|y,z,f(y,z))
\end{equation}

To prove the position and momentum relations requires the same sequence of properties, so we prove the position relation here.

To obtain the bound for $h(x_{A}|x_{B},x_{C})$, we use property (c):
\begin{equation}
h\Big(\sum_{i=A,B,C}\!\!\eta_{i}x_{i}\Big)\geq h\Big(\sum_{i=A,B,C}\!\!\eta_{i}x_{i}\Big|\sum_{j=B,C}\!\!\eta_{j}x_{j}\Big),
\end{equation}
and then property (d):
\begin{equation}
h\Big(\sum_{i=A,B,C}\!\!\eta_{i}x_{i}\Big|\sum_{j=B,C}\!\!\eta_{j}x_{j}\Big)=h\Big(\eta_{A}x_{A}\Big|\sum_{j=B,C}\!\!\eta_{j}x_{j}\Big)
\end{equation}
Next, we use property (c):
\begin{equation}
h\Big(\eta_{A}x_{A}\Big|\sum_{j=B,C}\!\!\eta_{j}x_{j}\Big)\geq h\Big(\eta_{A}x_{A}\Big|\sum_{j=B,C}\!\!\eta_{j}x_{j},x_{B},x_{C}\Big)
\end{equation}
followed by property (e):
\begin{equation}
h\Big(\eta_{A}x_{A}\Big|\sum_{j=B,C}\!\!\eta_{j}x_{j},x_{B},x_{C}\Big)=h(\eta_{A}x_{A}|x_{B},x_{C})
\end{equation}
and finally, we use property (a):
\begin{equation}
h(\eta_{A}x_{A}|x_{B},x_{C})= h(x_{A}|x_{B},x_{C})+\log(|\eta_{A}|)
\end{equation}
so that the bound becomes:
\begin{equation}
h\Big(\!\!\!\sum_{i=A,B,C}\!\!\eta_{i}x_{i}\Big)\geq h(x_{A}|x_{B},x_{C})+\log(|\eta_{A}|)
\end{equation}
Using the same sequence of properties, we can also derive:
\begin{align}
h\Big(\!\!\!\sum_{i=A,B,C}\!\!\eta_{i}x_{i}\Big)&\geq h(x_{B}|x_{A},x_{C})+\log(|\eta_{B}|)\\
h\Big(\!\!\!\sum_{i=A,B,C}\!\!\eta_{i}x_{i}\Big)&\geq h(x_{C}|x_{A},x_{B})+\log(|\eta_{C}|)
\end{align}
which proves the general position relation \eqref{posrel}.

\section{Derivation of triphoton wavefunction for cascaded SPDC}\label{ApD}
In this section, we show how to derive the triphoton spatial wavefunction in cascaded SPDC. To begin, we have the Hamiltonian for the SPDC process:
\begin{equation}
\hat{H}_{SPDC}=\sum_{k_{p},k_{1},k_{2}}G_{k_{p},k_{1},k_{2}}\hat{a}_{k_{p}}\hat{a}^{\dagger}_{k_{1}}\hat{a}^{\dagger}_{k_{2}} + h.c.
\end{equation}
To describe the cascaded SPDC process, we use first-order time-dependent perturbation theory to describe the evolution of the field after interacting with each crystal. After the first interaction $(k_{p}\rightarrow k_{1}+k_{2})$, and after the second interaction, $(k_{2}\rightarrow k_{3}+k_{4})$. The approximate state of the down-converted field after these interactions is:
\begin{align}
|\psi\rangle_{field}&\approx\left(\mathbf{I} + \sum_{k_{2'}k_{3}k_{4}}G_{k_{2'}k_{3}k_{4}}\hat{a}_{k_{2'}}\hat{a}^{\dagger}_{k_{3}}\hat{a}^{\dagger}_{k_{4}}\right)\nn\\
\cdot&\left(\mathbf{I} + \sum_{k_{p}k_{1}k_{2}}G_{k_{p}k_{1}k_{2}}\hat{a}_{k_{p}}\hat{a}^{\dagger}_{k_{1}}\hat{a}^{\dagger}_{k_{2}}\right)|vac\rangle
\end{align}
To first non-trivial order in the generation of photon triplets, the state of the photon triplets is described by:
\begin{align}
|\psi\rangle_{CDC}&\approx\!\!\!\!\sum_{\substack{k_{2'}k_{3}k_{4} \\k_{p}k_{1}k_{2}}}G_{k_{2'}k_{3}k_{4}}G_{k_{p}k_{1}k_{2}}\hat{a}_{k_{p}}\hat{a}_{k_{2'}}\hat{a}^{\dagger}_{k_{2}}\hat{a}^{\dagger}_{k_{1}}\hat{a}^{\dagger}_{k_{3}}\hat{a}^{\dagger}_{k_{4}}|vac\rangle\nn\\
&\!\!\!\!\!\!\!\!\!\!\!\!\!\!\!\!\!\!=\left(\sum_{k_{p}k_{1}k_{3}k_{4}}\left(\sum_{k_{2}}G_{k_{2}k_{3}k_{4}}G_{k_{p}k_{1}k_{2}}\right)\hat{a}_{k_{p}}\hat{a}^{\dagger}_{k_{1}}\hat{a}^{\dagger}_{k_{3}}\hat{a}^{\dagger}_{k_{4}}\right)|vac\rangle
\end{align}
where the simplification is carried out by the identity:
\begin{equation}
\hat{a}_{k_{2'}}\hat{a}_{k_{2}}^{\dagger}= \hat{a}_{k_{2}}^{\dagger}\hat{a}_{k_{2'}} + \delta_{k_{2},k_{2'}}
\end{equation}

In \cite{Schneeloch_SPDC_Reference_2016}, the spatially varying components of $G_{k_{p},k_{1},k_{2}}$ is given by:
\begin{equation}
G_{k_{p},k_{1},k_{2}}\propto \alpha_{k_{p}}\int d^{3}r \big(\chi_{\text{eff}}^{(2)}(\vec{r})e^{-i\vec{\Delta k}\cdot \vec{r}}\big)
\end{equation}
where $\chi_{eff}^{(2)}(\vec{r})$ is the spatially varying second order nonlinear susceptibility taken to be a constant inside the nonlinear medium unless performing quasi phase matching by periodic poling in which case it flips sign with the flipping poling. Then, for a rectangular crystal of length $L_{z}$ (and other dimensions $L_{x}$ and $L_{y}$), this integrates to:
\begin{align}
G_{k_{p},k_{1},k_{2}}&=\mathcal{N}\alpha_{k_{p}} \prod_{i=x,y,z}\text{sinc}\left(\frac{\Delta k_{i} L_{i}}{2}\right)\\
 \Delta k_{i}&= k_{1i} + k_{2i} - k_{pi} -k_{\Lambda i}
\end{align}
where $\mathcal{N}$ is a normalization constant; $k_{\Lambda i}$ is the $i$-th component of the poling momentum $2\pi / \Lambda$; and $\Lambda$ is the poling period. In the bulk crystal case without periodic poling, $k_{\Lambda}=0$.

With the approximate expression for $G_{k_{p},k_{1},k_{2}}$, and assuming the pump is bright enough to replace its annihilation operator with the corresponding coherent state amplitude, we can express the state of the cascaded down-converted light in terms of a triphoton amplitude:
\begin{equation}
|\psi\rangle_{CDC}\approx\sum_{k_{1}k_{3}k_{4}}\Psi(\vec{k}_{1},\vec{k}_{3},\vec{k_{4}})\hat{a}^{\dagger}_{k_{1}}\hat{a}^{\dagger}_{k_{3}}\hat{a}^{\dagger}_{k_{4}}|vac\rangle
\end{equation}
where
\begin{equation}
\Psi(\vec{k}_{1},\vec{k}_{3},\vec{k_{4}})=\sum_{k_{p}}\alpha_{k_{p}}\left(\sum_{k_{2}}G_{k_{2}k_{3}k_{4}}G_{k_{p}k_{1}k_{2}}\right)
\end{equation}

With the approximate expression for $G_{k_{p},k_{1},k_{2}}$, we approximate the sums in the triphoton amplitude as integrals and note the following simplifications. We assume the pump is sufficiently narrowband in frequency that its longitudinal momentum takes on one value in this sum. Next, we take the small-angle/paraxial approximation so that the pump amplitude $\alpha_{kp}$ factors into the product of a longitudinal amplitude $\alpha_{pz}$ and a transverse amplitude $\alpha_{qp}$. Together, this gives:
\begin{align}
\Psi&(\vec{k}_{1},\vec{k}_{3},\vec{k_{4}})\approx \mathcal{N}\!\alpha_{pz}\!\int dk_{px}dk_{py} \alpha_{qp}(\vec{q}_{p})\int\!\!\!\Big(\!\!\!\prod_{i=x,y,z}\!\!\!dk_{2i}\Big)\nn\\
&\cdot\Bigg(\prod_{i=x,y,z}\text{sinc}\Bigg(\frac{(k_{1i}+k_{2i}-k_{pi}-k_{\Lambda 1i}) L_{i}}{2}\Bigg)\cdot\nn\\
&\cdot\text{sinc}\Bigg(\frac{(k_{3i}+k_{4i}-k_{2i}-k_{\Lambda 2i}) L_{i}}{2}\Bigg)\Bigg)
\end{align}
where $\mathcal{N}$ is a normalization constant, and $\vec{q}_{p}$ is the projection of $\vec{k}_{p}$ onto the transverse $(xy)$ plane. Here, we are assuming that cascaded SPDC is achieved either simultaneously in the same crystal using two different poling momenta $\Lambda_{1}$ and $\Lambda_{2}$, or using a sequence of two identical crystals both of length $L_{z}$ for simplicity. This integral has the form of a convolution, and can be solved to give:
\begin{align}
&\Psi(\vec{k}_{1},\vec{k}_{3},\vec{k_{4}})\approx \mathcal{N}\int dk_{px}dk_{py} \alpha_{qp}(\vec{q}_{p})\nn\\
&\cdot\prod_{i=x,y,z}\text{sinc}\left(\frac{(k_{3i}+k_{4i}+k_{1i}- k_{pi}-k_{\Lambda 1i}-k_{\Lambda 2i}) L_{i}}{2}\right)
\end{align}
Through the rest of this derivation $\mathcal{N}$ will be a normalization constant absorbing factors not dependent on transverse momentum including the longitudinal pump amplitude. Next, we assume the transverse crystal dimensions $L_{x}$ and $L_{y}$ are large enough to wholly encompass the beam without clipping any side, which in turn is much larger than the pump wavelength. The transverse sinc functions can only contribute significantly for values less than the order of $2\pi/L_{x}$ or $2\pi/L_{y}$, which is multiple orders of magnitude smaller than the pump momentum $2\pi/\lambda_{p}$. Because of this, they can be treated as delta functions when integrating over the transverse components of the pump momentum. In addition, this enforces transverse momentum conservation.
\begin{align}
&\Psi(\vec{k}_{1},\vec{k}_{3},\vec{k_{4}})\approx \mathcal{N}\int dk_{px}dk_{py} \alpha_{qp}(\vec{q}_{p})\nn\\
&\cdot\text{sinc}\left(\frac{(k_{3z}+k_{4z}+k_{1z}- k_{pz}-k_{\Lambda 1z}-k_{\Lambda 2z}) L_{z}}{2}\right)\nn\\
&\cdot\prod_{i=x,y}\delta(k_{3i}+k_{4i}+k_{1i}- k_{pi})
\end{align}
For simplification, we will assume the periodic poling (when necessary) only exists in the longitudinal direction so that for $i=(x,y)$, $k_{\Lambda 1i}=k_{\Lambda 2i}=0$. 

At this point, we have a three-dimensional triphoton aplitude, describing both longitudinal and transverse components of the photons' momenta. To isolate the transverse spatial component of the triphoton amplitude, we express the longitudinal momentum components in the sinc function terms of the respective momentum magnitudes and the transverse spatial components through the Pythagorean formula:
\begin{equation}
k_{z} = \sqrt{|\vec{k}|^{2} - |\vec{q}|^{2}} \approx |\vec{k}|-\frac{|\vec{q}|^{2}}{2|\vec{k}|}
\end{equation}
The approximation comes from the small-angle approximation, in which the magnitude $|\vec{q}|$ is considered small relative to $|\vec{k}|$. Initially, this complicates the sinc function:
\begin{align}
&\Psi(\vec{k}_{1},\vec{k}_{3},\vec{k_{4}})\approx \mathcal{N}\int dk_{px}dk_{py} \alpha_{qp}(\vec{q}_{p})\nn\\
&\cdot\text{sinc}\Bigg(\frac{(|\vec{k}_{3}|+|\vec{k}_{4}|+|\vec{k}_{1}|- |\vec{k}_{p}|-|\vec{k}_{\Lambda 1}|-|\vec{k}_{\Lambda 2}|) L_{z}}{2}+\nn\\
&- \Big(\frac{|\vec{q}_{3}|^{2}}{|\vec{k}_{3}|}+\frac{|\vec{q}_{4}|^{2}}{|\vec{k}_{4}|}+\frac{|\vec{q}_{1}|^{2}}{|\vec{k}_{1}|}-\frac{|\vec{q}_{p}|^{2}}{|\vec{k}_{p}|}\Big)\frac{L_{z}}{4}\Bigg)\nn\\
&\cdot\prod_{i=x,y}\delta(k_{3i}+k_{4i}+k_{1i}- k_{pi}).
\end{align}

Next, we assume the use of narrowband frequency filters to fix the magnitudes $|\vec{k}_{1}|$, $|\vec{k}_{3}|$, and $|\vec{k}_{4}|$. The pump momentum magnitude $|\vec{k}_{p}|$ is already fixed by the prior assumption of a narrow linewidth laser, and the poling momenta are constants of the periodically poled crystals. Through an optimum choice of poling momenta, we can obtain the condition:
\begin{equation}
(|\vec{k}_{3}|+|\vec{k}_{4}|+|\vec{k}_{1}|- |\vec{k}_{p}|-|\vec{k}_{\Lambda 1}|-|\vec{k}_{\Lambda 2}|)\frac{L_{z}}{2}\approx 0
\end{equation}
greatly simplifying the triphoton spatial amplitude. With this, the rest of the triphoton amplitude approximately factors into the product of a longitudinal amplitude, and the transverse spatial amplitude we are seeking to calculate.

To isolate the quantum state describing just the transverse momenta of the triphoton, we would formally have to trace over their longitudinal momenta, which would generally result in a mixed state. To the extent that the triphoton ampltude does factor as the product of a transverse amplitude, and a longitudinal amplitude, the tracing over will produce a pure transverse amplitude, which can be obtained automatically by neglecting the longitudinal term of the triphoton amplitude and redefining the normalization constant accordingly:
\begin{align}
&\psi(\vec{q}_{1},\vec{q}_{3},\vec{q_{4}})\approx \mathcal{N}\int dk_{px}dk_{py} \alpha_{qp}(\vec{q}_{p})\nn\\
&\cdot \text{sinc}\Bigg(\Big(\frac{|\vec{q}_{3}|^{2}}{|\vec{k}_{3}|}+\frac{|\vec{q}_{4}|^{2}}{|\vec{k}_{4}|}+\frac{|\vec{q}_{1}|^{2}}{|\vec{k}_{1}|}-\frac{|\vec{q}_{p}|^{2}}{|\vec{k}_{p}|}\Big)\frac{L_{z}}{4}\Bigg)\nn\\
&\cdot\prod_{i=x,y}\delta(k_{3i}+k_{4i}+k_{1i}- k_{pi}).
\end{align}

By assuming the pump beam is well-collimated, its transverse momentum bandwidth determined by $\alpha_{p}(\vec{q}_{p})$ is sufficiently narrow (and centered at $\vec{q}_{p}=0$) that we can neglect the sinc function's dependence on $\vec{q}_{p}$:
\begin{align}
&\psi(\vec{q}_{1},\vec{q}_{3},\vec{q_{4}})\approx \mathcal{N}\int dk_{px}dk_{py} \alpha_{qp}(\vec{q}_{p})\nn\\
&\cdot\text{sinc}\left(\left(\frac{|\vec{q}_{3}|^{2}}{|\vec{k}_{3}|}+\frac{|\vec{q}_{4}|^{2}}{|\vec{k}_{4}|}+\frac{|\vec{q}_{1}|^{2}}{|\vec{k}_{1}|}\right)\frac{L_{z}}{4}\right)\nn\\
&\cdot\prod_{i=x,y}\delta(k_{3i}+k_{4i}+k_{1i}- k_{pi})
\end{align}
Finally, we can perform the integration over the transverse components of the pump momentum to obtain:
\begin{align}
&\psi(\vec{q}_{1},\vec{q}_{3},\vec{q_{4}})\approx \mathcal{N}\alpha_{qp}(\vec{q}_{1}+\vec{q}_{3}+\vec{q}_{4})\nn\\
&\cdot\text{sinc}\Bigg(\Bigg(\frac{|\vec{q}_{1}+\vec{q}_{4}-\vec{q}_{p}|^{2}}{|\vec{k}_{3}|}+\nn\\
&+\frac{|\vec{q}_{1}+\vec{q}_{3}-\vec{q}_{p}|^{2}}{|\vec{k}_{4}|}+\frac{|\vec{q}_{3}+\vec{q}_{4}-\vec{q}_{p}|^{2}}{|\vec{k}_{1}|}\Bigg)\frac{L_{z}}{4}\Bigg)
\end{align}
This expression is further simplified using the assumption we already made of the well-collimated (narrow transverse momentum bandwidth) pump, and one other. As before, the sinc function's dependence on $\vec{q}_{p}$ can be eliminated because it is nearly constant over the range of $\vec{q}_{p}$ determined by the transverse pump amplitude $\alpha_{qp}$:
\begin{align}
&\psi(\vec{q}_{1},\vec{q}_{3},\vec{q_{4}})\approx \mathcal{N}\alpha_{qp}(\vec{q}_{1}+\vec{q}_{3}+\vec{q}_{4})\nn\\
&\cdot\text{sinc}\Bigg(\Bigg(\frac{|\vec{q}_{1}+\vec{q}_{4}|^{2}}{|\vec{k}_{3}|}+\frac{|\vec{q}_{1}+\vec{q}_{3}|^{2}}{|\vec{k}_{4}|}+\frac{|\vec{q}_{3}+\vec{q}_{4}|^{2}}{|\vec{k}_{1}|}\Bigg)\frac{L_{z}}{4}\Bigg)
\end{align}

By assuming nearly collinear propagation, and fitting narrowband frequency filters to photon detectors, we can approximately fix the magnitudes of the down-converted photons' momenta $|\vec{k}_{1}|$, $|\vec{k}_{3}|$, and $|\vec{k}_{4}|$. For our triphoton wavefunction, we are considering degenerate collinear photon triplets generated from cascaded SPDC, so that $|\vec{k}_{1}|=|\vec{k}_{3}|=|\vec{k}_{4}|=k_{\tilde{p}}/3$. With this, we obtain the identical phase-matching function discussed in the body of the paper:
\begin{align}
&\psi(\vec{q}_{1},\vec{q}_{3},\vec{q}_{4})=\mathcal{N}\alpha_{qp}(\vec{q}_{1}+\vec{q}_{3}+\vec{q}_{4})\cdot\nn\\
&\cdot \text{sinc}\left(\frac{L_{z}}{4|k_{1}|}\left(|\vec{q}_{3}+\vec{q}_{4}|^{2}+|\vec{q}_{1}+\vec{q}_{4}|^{2}+|\vec{q}_{1}+\vec{q}_{3}|^{2}\right)\right)
\end{align}
The additional complication that arises because in general $|\vec{k}_{1}|\neq |\vec{k}_{3}|\neq |\vec{k}_{4}|$ due to the small variation of refractive index with propagation direction in the collinear regime increases the difficulty of computing the triphoton wavefunction without providing new insight into the magnitude and quality of correlations present. For that reason, we invoke this approximation $|\vec{k}_{1}|=|\vec{k}_{3}|=|\vec{k}_{4}|=k_{\tilde{p}}/3$ in the body of the paper.

\section{Exact tripartite entanglement and Schmidt decomposition of triple Gaussian wavefunction}\label{ApE}
In this appendix, we describe how to find the marginal eigenvalues of system $A$ of tri-partite system $ABC$ described by the triple-Gaussian wavefunction, as well as its exact Schmidt decomposition.

The double-Gaussian wavefunction $\psi(x_{A},x_{B})$ is given by:
\begin{equation}
\psi(x_{A},x_{B})=\frac{1}{\sqrt{2\pi\sigma_{+}\sigma_{-}}}e^{-\frac{(x_{A}+x_{B})^{2}}{8\sigma_{+}^{2}}}e^{-\frac{(x_{A}-x_{B})^{2}}{8\sigma_{-}^{2}}}
\end{equation}
This wavefunction has the following Schmidt decomposition:
\begin{equation}
\psi(x_{A},x_{B})=\sum_{n=0}^{\infty}\sqrt{\lambda_{n}}\phi_{n}(x_{A})\theta_{n}(x_{B})
\end{equation}
where
\begin{equation}
\lambda_{n}=\frac{4r}{(r+1)^2}\left(\frac{r-1}{r+1}\right)^{2n};
\end{equation}
$r$ is the ratio of $\sigma(x_{A}+x_{B})/\sigma(x_{A}-x_{B})$ which simplifies to $\sigma_{+}/\sigma_{-}$, and $\phi_{n}(x)$ is the $n$th-order Hermite-Gaussian wavefunction as seen in the solutions to the quantum harmonic oscillator. For the double-Gaussian wavefunction, $\theta_{n}(x)=\phi_{n}(x)$, though for arbitrary wavefunctions, they can be different.

If we take the marginal density operator of the double-Gaussian state, we find that:
\begin{equation}
\hat{\rho}_{A}=\int dx_{A}dx_{A}'\eta(x_{A},x_{A}')|x_{A}\rangle\langle x_{A}'|
\end{equation}
where
\begin{equation}
\eta(x_{A},x_{A}')=\sum_{n}\lambda_{n}\phi_{n}(x_{A})\phi_{n}(x_{A}')
\end{equation}

What is important to note are two points. First is that the Schmidt eigenvalues $\lambda_{n}$ are geometrically distributed, and second, that the ratio between successive Schmidt coefficients $\sqrt{\lambda_{n}}$ is given by:
\begin{equation}
\frac{\sqrt{\lambda_{n+1}}}{\sqrt{\lambda_{n}}}=\frac{r-1}{r+1}.
\end{equation}

\subsection{The reduced density operator for the triple-Gaussian wavefunction}
The Triple-Gaussian state $|\psi\rangle_{ABC}$ is given by:
\begin{align}
|\psi\rangle_{ABC}&=\int dx_{A}dx_{B}dx_{C}|x_{A},x_{B},x_{C}\rangle \langle x_{A},x_{B},x_{C}|\psi\rangle_{ABC}\\
&=\int dx_{A}dx_{B}dx_{C} |x_{A},x_{B},x_{C}\rangle\psi(x_{A},x_{B},x_{C}) 
\end{align}
where $\psi(x_{A},x_{B},x_{C})$ is the triple-Gaussian wavefunction. 
\begin{equation}
\psi(x_{A},x_{B},x_{C}) =\frac{e^{-\frac{(x_{A}+x_{B}+x_{C})^{2}}{12\sigma_{u}^{2}}- \frac{\left(x_{A}-\frac{x_{B}+x_{C}}{2}\right)^{2}}{6\sigma_{v}^{2}}-\frac{(x_{B}-x_{C})^{2}}{8\sigma_{w}^{2}}}}{\sqrt{(2\pi)^{3/2} \sigma_{u}\sigma_{v}\sigma_{w}}}
\end{equation}
where $\sigma_{w}$ is set equal to $\sigma_{v}$ for rotational symmetry around the $x_{u}$-axis.

The density matrix $\hat{\rho}_{ABC}$ corresponding to this state is $|\psi\rangle_{ABC}\langle \psi|_{ABC}$:
\begin{align}
\hat{\rho}_{ABC}
=&\int dx_{A}dx_{A}'dx_{B}dx_{B}'dx_{C}dx_{C}'\nn\\
&\quad\quad\cdot|x_{A},x_{B},x_{C}\rangle\langle x_{A}',x_{B}',x_{C}'|\nn\\
&\quad\quad\quad\quad\cdot\psi(x_{A},x_{B},x_{C})\psi^{*}(x_{A}',x_{B}',x_{C}')
\end{align}
Next, we obtain $\hat{\rho}_{A}$ by tracing over $B$ and $C$ so that we may later obtain the marginal eigenvalues:
\begin{align}
\hat{\rho}_{A}&=\text{Tr}_{BC}[\hat{\rho}_{ABC}]\\
&=\int d\mu_{B} d\mu_{C}\langle\mu_{B},\mu_{C}|\hat{\rho}_{ABC}|\mu_{B},\mu_{C}\rangle\\
&=\int dx_{A}dx_{A}'|x_{A}\rangle\langle x_{A}'|\;\eta(x_{A},x_{A}')
\end{align}
where
\begin{equation}
\eta(x_{A},x_{A}')=\int d\mu_{B} d\mu_{C}
\psi(x_{A},\mu_{B},\mu_{C})\psi^{*}(x_{A}',\mu_{B},\mu_{C})
\end{equation}
Here, $\eta(x_{A},x_{A}')$ is the density operator function describing $\hat{\rho}_{A}$.

Now, for the triple-Gaussian wavefunction, $\eta(x_{A},x_{A}')$ has a double-Gaussian form:
\begin{align}
\eta(x_{A},x_{A}')= &\sqrt{\frac{3}{2\pi(\sigma_{u}^{2} + 2\sigma_{v}^{2})}}e^{-\frac{(x_{A}-x_{A}')^{2}}{24}(\frac{1}{\sigma_{u}^{2}} +\frac{2}{\sigma_{v}^{2}})}\nn\\
&\cdot e^{-\frac{9(x_{A}+x_{A}')^{2}}{24}(\frac{1}{\sigma_{u}^{2} + 2\sigma_{v}^{2}})}
\end{align}
Since this function is up to a normalization constant, equal to a double-Gaussian wavefunction, we can define its correlation ratio $R$ as $\sigma(x_{A}+x_{A}')/\sigma(x_{A}-x_{A}')$, and obtain the result:
\begin{equation}\label{schmidtR}
R= \frac{1}{3}\sqrt{5+\frac{2\sigma_{u}^{4} + 2\sigma_{v}^{4}}{\sigma_{u}^{2}\sigma_{v}^{2}}}
\end{equation}

\subsection{Determining marginal eigenvalues through analogy}
Importantly, $\eta(x_{A},x_{A}')$ admits precisely the same type of Schmidt decomposition as the double-Gaussian wavefunction:
\begin{equation}
\eta(x_{A},x_{A}')=\sum_{n=0}^{\infty}G\sqrt{\nu_{n}}\phi_{n}(x_{A})\theta_{n}(x_{A}')
\end{equation}
where $G$ is a normalization constant. Here again, $\theta_{n}(x)=\phi_{n}(x)$.

Because of this, $\eta(x_{A},x_{A}')$ for the triple-Gaussian, is mathematically identical to $\eta(x_{A},x_{A}')$ for the double-Gaussian. In particular, we deduce that the eigenvalues of $\hat{\rho}_{A}$ are given by:
\begin{equation}
\lambda_{n}=G\sqrt{\nu_{n}}
\end{equation}

Then, the ratio of successive Schmidt coefficients is given by:
\begin{equation}
\frac{\sqrt{\nu_{n+1}}}{\sqrt{\nu_{n}}}=\frac{\lambda_{n+1}}{\lambda_{n}}=\frac{R-1}{R+1}
\end{equation}
With this set of ratios, we can normalize and find the total list of marginal eigenvalues of $\hat{\rho}_{A}$:
\begin{equation}\label{schmidteigs}
\lambda_{n}=\frac{2}{1+R}\Big(\frac{R-1}{R+1}\Big)^{n}
\end{equation}
where $R$ is given in \eqref{schmidtR}.

With this eigenvalue distribution (i.e., the geometric distribution) the von Neumann entropy $S(A)$ has the simple form:
\begin{equation}
S(A)=\frac{h_{2}(\lambda_{0})}{\lambda_{0}}
\end{equation}
where $h_{2}(\lambda)$ is the binary entropy function:
\begin{equation}
h_{2}(\lambda)=-\lambda\log_{2}(\lambda) -(1-\lambda)\log_{2}(1-\lambda).
\end{equation}
Due to the triple Gaussian wavefunction referring to a pure state, and one that is symmetric between parties, this von Neumann entropy is also equal to the tripartite entanglement of formation of this state.

\subsubsection{On the Schmidt decomposition of triphoton wavefunction:}
With the marginal eigenvalues of the triple-Gaussian wavefunction known, it is tempting to alledge that one can Schmidt-decompose this wavefunction into a series of triplets of Hermite-Gaussian modes, similar to the case for the double-Gaussian wavefunction:
\begin{equation}
\psi(x_{A},x_{B},x_{C})\neq\sum_{n}\sqrt{\lambda_{n}}\phi_{n}(x_{A})\phi_{n}(x_{B})\phi_{n}(x_{C})
\end{equation}
This decomposition is not valid because the triple-Gaussian waveunction for $x_{C}=0$ reduces to a double-Gaussian wavefunction, but the alleged decomposition cannot because $\phi_{n}(0)$ is not a geometric series in $n$. 

However, we find that the true Schmidt decomposition of $\psi(x_{A},x_{B},x_{C})$ (for a bipartite split, e.g., $A\otimes BC$) still involves Hermite-Gaussian wavefunctions, and is of the following form:
\begin{equation}
\psi(x_{A},x_{B},x_{C})=\sum_{n}\sqrt{\lambda_{n}}\phi_{n}(x_{A})\theta_{n}(x_{B},x_{C})
\end{equation}
where $\phi_{n}(x_{A})$ is the $n$-th order Hermite-Gaussian wavefunction:
\begin{equation}
\phi_{n}(x_{A}) = \frac{1}{(2\pi\sigma_{A}^{2})^{1/4}}\frac{1}{\sqrt{n!2^{n}}}H_{n}\left(\frac{x}{\sigma_{A}\sqrt{2}}\right)e^{-\frac{x_{A}^{2}}{4\sigma_{A}^{2}}}
\end{equation}
where $H_{n}(x)$ is the $n$-th order Hermite polynomial of $x$, and $\sigma_{A}$ is given as:
\begin{equation}\label{sigmaA}
\sigma_{A}=\sqrt{\sigma_{u}\sigma_{v}}\left(\frac{\sigma_{u}^{2} + 2 \sigma_{v}^{2}}{2\sigma_{u}^{2} + \sigma_{v}^{2}}\right)^{1/4}
\end{equation}
$\sigma_{A}$ is not to be confused with $\sigma(x_{A})$, the standard deviation of the marginal position, which is $\sqrt{(\sigma_{u}^{2} + 2\sigma_{v}^{2})/3}$.

In this decomposition, $\theta_{n}(x_{B},x_{C})$ is an $n$-th order polynomial double-Gaussian wavefunction, defined here as the product of a Gaussian wavefunction in the rotated coordinate $(x_{B}-x_{C})$ and an $n$-th order polynomial Gaussian wavefunction in the orthogonal coordinate $(x_{B}+x_{C})$. The explicit form is given by the integral:
\begin{equation}
\sqrt{\lambda_{n}}\theta_{n}(x_{B},x_{C})=\int dx_{A} \phi_{n}^{*}(x_{A})\psi(x_{A},x_{B},x_{C})
\end{equation}
and $\theta_{n}(x_{B},x_{C})$ is determined through renormalization. From this, we were able to determine $\sigma_{A}$ \eqref{sigmaA}, because only the correct value of $\sigma_{A}$ would yield the right eigenvalue spectrum $\lambda_{n}$ matching with what we obtained with our previous analogy method \eqref{schmidteigs}.
 
To simplify the integral, $\psi(x_{A},x_{B},x_{C})$ can be re-expressed as:
\begin{equation}
\psi(x_{A},x_{B},x_{C}) =\frac{e^{-\frac{(x_{A}+\sqrt{2}x_{p})^{2}}{12\sigma_{u}^{2}}- \frac{\left(x_{A}-\frac{x_{p}}{\sqrt{2}}\right)^{2}}{6\sigma_{v}^{2}}-\frac{x_{m}^{2}}{4\sigma_{v}^{2}}}}{\sqrt{(2\pi)^{3/2} \sigma_{u}\sigma_{v}^{2}}}
\end{equation}
such that:
\begin{equation}
x_{p}=\frac{x_{B}+x_{C}}{\sqrt{2}}\qquad:\qquad x_{m}=\frac{x_{B}-x_{C}}{\sqrt{2}}
\end{equation}
Then, because $\psi(x_{A},x_{B},x_{C})$ factors into a product of a function $g(x_{A},x_{p})$ and a Gaussian $f(x_{m})$, the integral simplifies:
\begin{equation}
\sqrt{\lambda_{n}}\theta_{n}(x_{B},x_{C})=f(x_{m})\int dx_{A} \phi_{n}^{*}(x_{A})g(x_{A},x_{p})
\end{equation}
Similar to Hermite-Gaussian wavefunctions, this simplified integral  is equal to a constant (varying with $n$) Gaussian wavefunction of $x_{p}$ times an $n$-th order polynomial of $x_{p}$. The form is not recognized by us, so we leave it as an open problem for the interested reader.

Let $\sigma_{u}=\gamma\sigma_{v}$; let $s=(\gamma^{2}-1) x_{p}\sigma_{A}$; and let $u=(1+2\gamma^{2}\sigma_{A}^{4} - 9\gamma^{4}\sigma_{v}^{4})$. Then we find:
\begin{align}
((1&+2\gamma^{2})\sigma_{A}^{2} + 3\gamma^{2}\sigma_{v}^{2})^{(n+1/2)}\left(\frac{(2\pi)^{(1/4)}}{2}\right)\cdot\nn\\
&\cdot \int dx_{A} g(x_{A},x_{p})\phi_{n}(x_{A})=\nn\\
&=e^{-x_{p}^{2}\left(\frac{3\sigma_{A}^{2} + (2+\gamma^{2})\sigma_{v}^{2}}{(1+2\gamma^{2})\sigma_{A}^{2} +3\gamma^{2}\sigma_{v}^{2}}\right)}P_{n}(s,u)
\end{align}
where up to order $n=7$ we have:
\begin{align}
P_{0}&=\sqrt{6}\nn\\
P_{1}&=2\sqrt{3} s \nn\\
P_{2}&=\sqrt{3}(2s^{2}-u)\nn\\
P_{3}&=\sqrt{2}s(2s^{2}-3u)\nn\\
P_{4}&=\frac{1}{2}\left(4s^{4}-12s^{2}u +3u^{2}\right)\nn\\
P_{5}&=\frac{s}{\sqrt{10}}(4s^{4} -20s^{2}u + 15u^{2})\nn\\
P_{6}&=\frac{1}{2\sqrt{30}}(8s^{6} - 60s^{4}u + 90 s^{2}u^{2}-15u^{3})\nn\\
P_{7}&=\frac{s}{2\sqrt{105}}(8s^{6}-84s^{2}u+210s^{2}u^{2}-105u^{3})
\end{align}

%

\end{document}